\documentclass[a4paper,10pt]{article}  

\usepackage{graphicx}

\usepackage{amsmath}
\usepackage{amssymb}
 
\usepackage[abbr]{harvard}
\citationstyle{dcu}
\citationmode{abbr}
\renewcommand{\harvardand}{\&}

\newcommand{\EXP}[1]{\mathrm{e}^{#1}} 
\newcommand{\DEF}{\stackrel{\mathrm{def}}{=}}
\newcommand{\DEFt}{\smash{\overset{\text{\tiny def}}{=}}}
\newcommand{\imat}{{\mathrm{i}}} 
\newcommand{\dmat}{\mathrm{d}}
\newcommand{\Omat}{\mathrm{O}}

\newcommand{\kp}{{\substack{k \\[-.7ex]\scriptscriptstyle \rightarrow}}}
\newcommand{\km}{{\substack{k \\[-.7ex]\scriptscriptstyle  \leftarrow}}}
\newcommand{\mkm}{{\substack{-k \\[-.7ex] \leftarrow}}}
\newcommand{\kpu}{{\substack{k_1 \\[-.5ex]\scriptscriptstyle \rightarrow}}}
\newcommand{\kpd}{{\substack{k_2 \\[-.5ex]\scriptscriptstyle \rightarrow}}}
\newcommand{\kmd}{{\substack{k_2 \\[-.5ex]\scriptscriptstyle \leftarrow}}}

\newcommand{\Poisson}[2]{\left\{\mkern-7.1mu\left\{\smash{#1,#2}\right\}\mkern-7.1mu\right\}}

\newcommand{\RR}{\mathbb{R}}
\newcommand{\ZZ}{\mathbb{Z}}
\newcommand{\dst}{\displaystyle}
 
\newcommand{\kete}[1]{|\kern.3ex#1\kern.3ex\rangle}

\newcommand{\brae}[1]{\langle\kern.3ex #1 \kern.3ex|} 
\begin{document}

\title{Interaction with a field: a simple integrable model 
with backreaction}
\author{Amaury Mouchet\\ \\
Laboratoire de Math\'ematiques 
  et Physique Th\'eorique,\\ Universit\'e Fran\c{c}ois Rabelais de Tours 
--- \textsc{\textsc{cnrs (umr 6083)}},\\
F\'ed\'eration Denis Poisson,\\
 Parc de Grandmont 37200
  Tours,  France. \\{mouchet@phys.univ-tours.fr}}

\date{\today}

\maketitle 

{PACS: 46.40.Cd, 
46.40.Ff, 
42.50.Ct, 	
03.65.Yz. 	
}

\begin{abstract}The classical model of an oscillator linearly coupled to a string 
captures, for a low price in technique, many general features of more
realistic models for describing a particle interacting with a field or
an atom in a electromagnetic cavity. The scattering matrix and the
asymptotic in and out waves on the string can be computed exactly and
the phenomenon of resonant scattering can be introduced in the
simplest way.  The dissipation induced by the coupling of the
oscillator to the string can be studied completely. In the case of a
d'Alembert string, the backreaction leads to an
Abraham-Lorentz-Dirac-like equation. In the case of a Klein-Gordon
string, one can see explicitely how radiation governs the
(meta)stability of the (quasi)bounded mode.
\end{abstract}

\section{Introduction}

\begin{figure}[!ht]
\center
\includegraphics[width=9cm]{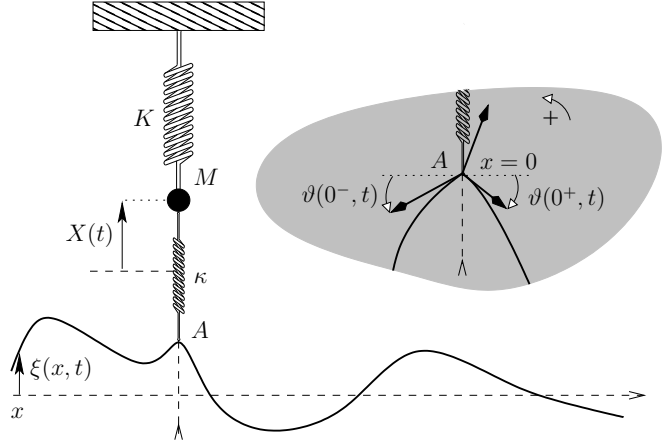}
\caption{\label{fig:cordosc}Model of a harmonic oscillator coupled to a string
via a massless spring. The long-dashed lines refer
to the equilibrium positions of the string and the oscillator. The inset shows the three forces that cancel
at the massless attachment point~$A$ ($x=0$): the left and right tensions 
of the string and the elastic force of the coupling spring. }
\end{figure}

Getting back to the spirit of the \textsc{xix}${}^\mathrm{th}$ century
--- when purely mechanical models were ubiquitous even for
understanding systems involving electromagnetic fields --- this paper
discusses a simple model of an oscillator coupled to a string that
presents a host of interesting features in a very accessible way. It
will be presented in detail in~\S~\ref{sec:model} (see
figure~\ref{fig:cordosc}). Even though the required 
technical background is
on an upper undergraduate/early graduate
 level, if one remains within a
classical (\textit{i.e.} non-quantum) context, this model encapsulates many
relevant features of physical interest which can be found in more
realistic models, for instance when considering the interaction of an
atom (the oscillator) with light (the string). 

First (\S~\ref{sec:scattering}), when the string is infinite, it
allows one to gain insight into (or to discover for the first time in
academic studies) the scattering of waves by a dynamical system.  It
provides the opportunity to introduce some of the ingredients of
scattering theory: when dealing with a one-dimensional mechanical
system, the notion of asymptotic states (the so-called ``in'' and
``out'' states) is made as simple as possible and the $S$ matrix takes
a particular simple form.  Unlike the transmission of light in a
Fabry-Perot interferometer, the resonant scattering appears here
explicitly in connection with the familiar resonance phenomenon of a
forced oscillator and we can also understand how the motion of the
oscillator affects in return the external excitation.

Second (\S~\ref{sec:radiation}), from the point of view 
of the oscillator, this model 
constitutes the most elementary example of radiation.
It shows concretely how an interaction not only induces a 
shift in the natural frequency of the oscillator (this can already 
be seen when only two degrees of freedom are coupled)  but also
that the coupling 
 to a large number of degrees of freedom 
(the string being seen as a large collection of oscillators)
induces a friction term  for the oscillator, although no dissipation exists
in the system as a whole \citeaffixed{Feynman+70a,KrivineLesne03a}{for an electric analogous phenomenon
 see \S~22.6 of}. Indeed, as far as I know, there are very few places 
where this model is discussed and always in the 
specialised literature with a d'Alembert string \cite{SollfreyGoertzel51a,Dekker85a}.
  Some of its variants \cite{Stevens61a,Rubin63a,Yurke84a,Dekker84a,Yurke86a}
are introduced precisely for studying dissipation at a quantum level. 
As we will see, a Klein-Gordon string allows to keep one discrete mode (a bounded state) without
dissolving it in the continuous spectrum of the string and therefore allows to mimic
the interaction of a field with a stable particle, not just a metastable one.
This model is particularly relevant to see how backreaction works: 
 For instance, for a d'Alembert string, the oscillator
is governed by an Abraham-Lorentz-Dirac-like equation.  Besides, 
we will be able to illustrate precisely the deep connection between the resonance
and the poles of the $S$ matrix (a major feature in high energy
particle physics and in condensed matter physics).

Before we give some guidelines for further developments in the conclusion (\S~\ref{sec:conclusion}),
we will complete our classical study in \S~\ref{sec:hamdiag} by the detailed diagonalization of the Hamiltonian
and the discussion of the completeness of the basis of modes that are used to described the dynamics.
This  will be the occasion to sketch the finite size effects if one wants to use this model for
describing an atom placed in a cavity. This study also prepares 
 the ground for the quantization
which will be proposed in a future paper.   

\section{The spring-string model}\label{sec:model}

\subsection{Description of the model}
A very thin homogeneous string of linear mass~$\mu_0$ is considered 
to have only transverse displacements in one direction. At equilibrium
it forms a straight line  along the $x$-axis with the uniform tension~$T_0$ 
(we will never take into account the effects of gravity).
The string is coupled to a system of mass~$M$
 with one degree
of freedom connected to a fixed support with a massless
spring of stiffness~$K$ (see figure~\ref{fig:cordosc}).
 The coupling is modelised by a second 
 massless spring of stiffness~$\kappa$ attached to the
string at the massless point~$A$ at~$x=0$.
 All the vibrations will be considered 
within the harmonic approximation of small amplitudes. We will 
denote by~$\xi(x,t)$ the transverse displacement at time~$t$ of the 
string element located at~$x$ at equilibrium. The displacement of~$M$
with respect to its equilibrium position will be denoted by~$X(t)$.

\subsection{Equations of motion}
 The equation of 
motion of the oscillator is
\begin{subequations}
\begin{equation}\label{eq:motionM}
M\frac{\dmat^2 X}{\dmat t^2}+ M\Omega^2_0\,X=-\kappa(X-\xi_0)
\end{equation}
where $\xi_0(t)\DEFt\xi(0,t)$  is the displacement of~$A$ and
$\Omega_0\DEFt\sqrt{K/M}$ is the harmonic frequency of the 
free oscillator.
For~$x\mathop{\neq}0$, $\xi$ fulfills the  one-dimensional d'Alem\-bert equation
\begin{equation}\label{eq:dalembert}
  \frac{\partial^2\xi}{\partial x^2}
 -\frac{1}{c^2}\frac{\partial^2\xi}{\partial t^2}=0
\end{equation}
 where the wave velocity on the string is given by~$c\DEFt\sqrt{T_0/\mu_0}$.
The derivation of~\eqref{eq:dalembert} is a major step in
every introductory course on 
waves \cite[\S~2.2, for instance]{Crawford68a}.
\begin{figure}[!ht]
\center
\includegraphics[width=8.5cm]{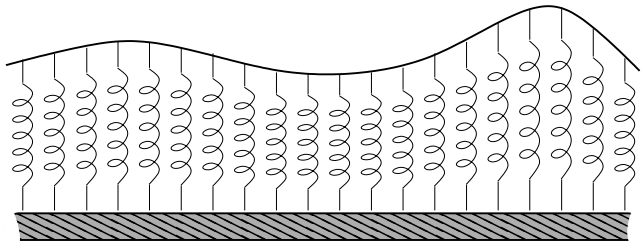}
\caption{\label{fig:cordematelas1}A mechanical model that 
leads to a Klein-Gordon equation: the string is attached to an
 elastic support that can be seen as a tight collection
of massless springs uniformly  distributed per unit length along the string. 
For another mechanical example,
see \protect\cite[\S~3.5]{Crawford68a}.}
\end{figure}
 Less common is perhaps the refinement of sticking the string to
 a ``mattress'' 
(see figure~\ref{fig:cordematelas1}) made of~$n$ massless springs 
per unit length along the~$x$-axis, each of them having a 
stiffness~$\varkappa$. 
The restoring force per unit length due to the mattress, $-n\varkappa\xi$,  
turns the d'Alembert equation into a Klein-Gordon equation  
\begin{equation}\label{eq:kleingordon}\tag{1b'}
 \frac{\partial^2\xi}{\partial x^2}
 -\frac{1}{c^2}\frac{\partial^2\xi}{\partial t^2}-\frac{\omega_0^2}{c^2}\,\xi=0\;,
\qquad  (x\neq0)
\end{equation}
where $\omega_0\DEFt c\sqrt{n\varkappa/T_0}$. This equation governs also the
propagation of the electromagnetic field in a rectangular waveguide \cite[chap.~24]{Feynman+70a}
and is also the relativistic equation of a quantum  particle
of mass~$\hbar\omega_0/c^2$, $c$ being the velocity of light in the vacuum.

Since the attachment point~$A$ is massless, the sum of the three
 forces applied to it 
vanishes as depicted in the inset of figure~\ref{fig:cordosc}. To first
order in the slope~$\vartheta(x,t)=\partial\xi/\partial x(x,t)$,
 the transverse component of the right tension applied to~$A$
 is given by~$T_0\vartheta(0^+,t)$ the limit when~$x\to0$ keeping~$x>0$.
On the other side, the transverse component of the left tension 
is~$-T_0\vartheta(0^-,t)$. The restoring force of the coupling string
corresponds to the opposite of the left hand side of 
\eqref{eq:motionM}, namely~$\kappa(X-\xi_0)$. Therefore the coupling
introduces a discontinuity of the slope of the string at~$x=0$:
\begin{equation}\label{eq:discontinuityslope}
  T_0\frac{\partial\xi}{\partial x}(0^+,t)
 -T_0\frac{\partial\xi}{\partial x}(0^-,t)=
  -\kappa\big(X(t)-\xi_0(t)\big)\;.
\end{equation}
\end{subequations}

\subsection{Dimensionless quantities}

The fundamental units will be chosen to be
 $T\DEFt M/(\mu_0c)$ for the times,
$L\DEFt M/\mu_0$ for the lengths and $M$ for the masses.
  The model is therefore
uniquely determined in terms of dimensionless quantities defined by
 the appropriate rescaling: $x_{\mathrm{eff}}\DEFt{x/L}$, 
$t_{\mathrm{eff}}\DEFt{t/T}$,
$\xi_{\mathrm{eff}}\DEFt\xi/L$, 
$X_{\mathrm{eff}}\DEFt X/L$, 
$\Omega_{0,\mathrm{eff}}\DEFt\Omega_0/(T^{-1})$,
$\kappa_{\mathrm{eff}}\DEFt\kappa/(MT^{-2})$, etc. These
 conventions correspond to 
$c_{\mathrm{eff}}\DEFt c/(LT^{-1})=1$. There is no need for any quantization 
as long as the effective Planck constant~$\hbar_{\mathrm{eff}}\DEFt\hbar/(ML^{2}T^{-1})=\hbar\mu_0/(M^2c)\ll1$.
 For simplifying the notations,
in the following we will drop the ``effective'' subscript and work directly 
with~$M\mathop{=}1$, $c\mathop{=}1$, $\mu_0\mathop{=}1$ and $T_0\mathop{=}1$.
Introducing the shifted\footnote{With more pedantry, one could speak of 
the renormalized frequency. Here the shift can be understood by the effective restoring force~$-(K+\kappa)X$ 
that~$M$ actually feels.} frequency
\begin{equation}
  \Omega_\kappa\DEF\sqrt{\Omega_0^2+\kappa}\;,
\end{equation}
the equations governing the dynamics
of the system become
\begin{subequations}\label{subeq:motiona}
\begin{equation}\label{eq:motionMa}
\frac{\dmat^2 X}{\dmat t^2}+\Omega^2_\kappa\,X=\kappa\,\xi_0\;,
\end{equation}
\begin{equation}\label{eq:kleingordona}
 \frac{\partial^2\xi}{\partial x^2}
 -\frac{\partial^2\xi}{\partial t^2}-\omega_0^2\,\xi=0\;, \qquad(x\neq0)
\end{equation}
and
\begin{equation}\label{eq:discontinuityslopea}
  \frac{\partial\xi}{\partial x}(0^+,t)
 -\frac{\partial\xi}{\partial x}(0^-,t)=
  \kappa\big(\xi_0(t)-X(t)\big)\;.
\end{equation}
with the help of the Dirac distribution~$\delta$, equations~\eqref{eq:kleingordona} and~\eqref{eq:discontinuityslopea} can be combined in one equation, valid for all~$x$,
\begin{equation}\label{eq:kleingordon2}\tag{3bc}
 \frac{\partial^2\xi}{\partial x^2}
 -\frac{\partial^2\xi}{\partial t^2}-\omega_0^2\,\xi=
\kappa\,\delta(x)\,(\xi-X)\;.
\end{equation}
Indeed, \eqref{eq:discontinuityslopea} is recovered
 after integrating \eqref{eq:kleingordon2} between~$x\mathop{=}-\epsilon$
 and~$x\mathop{=}+\epsilon$ when~$\epsilon\to0^+$, since~$\xi$ and its time derivatives
are continuous (and finite) everywhere.  
\end{subequations}

\subsection{The Hamiltonian}\label{subec:hamiltonian}

If one wants to prepare the ground for some  perturbative treatment of 
some non-linear corrections, if one wants 
to quantize the model and/or to couple it to a thermal bath, 
 one possible starting point is 
the Hamiltonian of the system expressed in term of
some canonical variables. The continuous part of the system (the string)
corresponds to a Hamiltonian density \cite[\S~12-4]{Goldstein80a} 
involving a pair of canonically 
conjugate fields~$\big(\pi(x,t),\xi(x,t)\big)$ whereas the oscillator
is described in terms of~$(P_X,X)$. The Poisson bracket  has also 
a mixed structure of continuous and discrete variables: For any two~$O_1$,
$O_2$ that are functions of~$(P_X,X)$ and functionals of~$(\pi,\xi)$,
\begin{equation}
  \Poisson{O_1}{O_2}\ \DEF\ \frac{\partial O_1}{\partial P_X}\frac{\partial O_2}{\partial X}
-\frac{\partial O_2}{\partial P_X}\frac{\partial O_1}{\partial X}+
\int 
\left(\frac{\delta O_1}{\delta \pi}\frac{\delta O_2}{\delta \xi}
-\frac{\delta O_2}{\delta \pi}\frac{\delta O_1}{\delta \xi}\right)\dmat x\;.
\end{equation} 
The Hamiltonian of the whole system generates the evolution of any 
function(al)~$O$ via~$\dmat O/\dmat t =\partial_tO+\Poisson{H}{O}$. For our system
we have
\begin{equation}\label{def:Ham}
  H\ \DEF\ \frac{1}{2}P_X^2+\frac{1}{2}\Omega_0^2X^2
           +\frac{1}{2}\kappa(X-\xi_0)^2
+\frac{1}{2}\int\bigg[\pi^2
                       +\left(\frac{\partial\xi}{\partial x}\right)^2
                       +\omega_0^2\,\xi^2
     \bigg]\dmat x\;;
\end{equation}
the corresponding Hamilton equations yield directly to \eqref{subeq:motiona}.
The interaction term is completely different from the one used in a recent one-dimensional
pedagogical model~\cite{Boozer07a}. The latter emphasizes
the recoil effect of the field on the mass~$M$ (unlike in the present work,
the external (translational) degrees of freedom for $M$ are considered)
whereas we are more interested in the resonance effects.

\section{Scattering}\label{sec:scattering}

\subsection{Free modes}

As for any vibrating system, a normal mode is defined to be
a particular collective motion where all the degrees of freedom oscillate
with the same frequency. In the absence of coupling~($\kappa=0$), 
for a given frequency~$\omega\geqslant0$, one can choose 
two independent normal  modes (the free modes)
 on the string  given 
by
\begin{equation}\label{eq:freemode}
  \xi^{\mathrm{fr}}_{\pm k}(x,t)=\frac{1}{\sqrt{2\pi}}
  \,\EXP{\imat(\pm kx-\omega t)}.
\end{equation}
The wave number is obtained from the 
dispersion relation of the Klein-Gordon equation:
\begin{equation}\label{eq:dispersion}
  k(\omega)\DEF\sqrt{\omega^2-\omega_0^2}\quad \Longleftrightarrow\quad\omega(k)=\sqrt{\omega_0^2+k^2}\;.
\end{equation}
When~$\omega<\omega_0$, $k$ is purely imaginary with a positive 
imaginary part and does not correspond to any traveling wave. When
$\omega>\omega_0$, $k$ is real positive and
 the two modes are two monochromatic 
counter-propagating waves. 

A real field~$\xi$ obeying the Klein-Gordon equation~\eqref{eq:kleingordona} carries an  linear energy
 density
$  \rho_e=\frac{1}{2}[({\partial\xi}/{\partial t})^2
                       +({\partial\xi}/{\partial x})^2
                       +\omega_0^2\xi^2
                ]
$ and a linear density of energy current $ j_e=-{\partial\xi}/{\partial t}\,{\partial\xi}/{\partial x}$.
 The conservation of energy takes the form~${\partial \rho_e}/{\partial t}+{\partial j_e}/{\partial x}=0$ for~$x\neq0$.
For a monochromatic traveling wave of complex amplitude~$a$, $\xi(x,t)=a\EXP{\imat(\pm kx-\omega t)}$, 
an elementary calculation shows that the average current over one period is given by
\begin{equation}\label{eq:je}
  \langle j_e\rangle=\pm k\omega|a|^2/2\qquad (k\ \text{real}),
\end{equation}
whereas for an evanescent wave
\begin{equation}\label{eq:je_ev}
  \langle j_e\rangle=0\qquad (k\ \text{imaginary}).
\end{equation}

\subsection{Definitions of the in and out asymptotic modes}\label{subsec:inoutdef}
\begin{figure}[!Ht]
\center
\includegraphics[width=\textwidth]{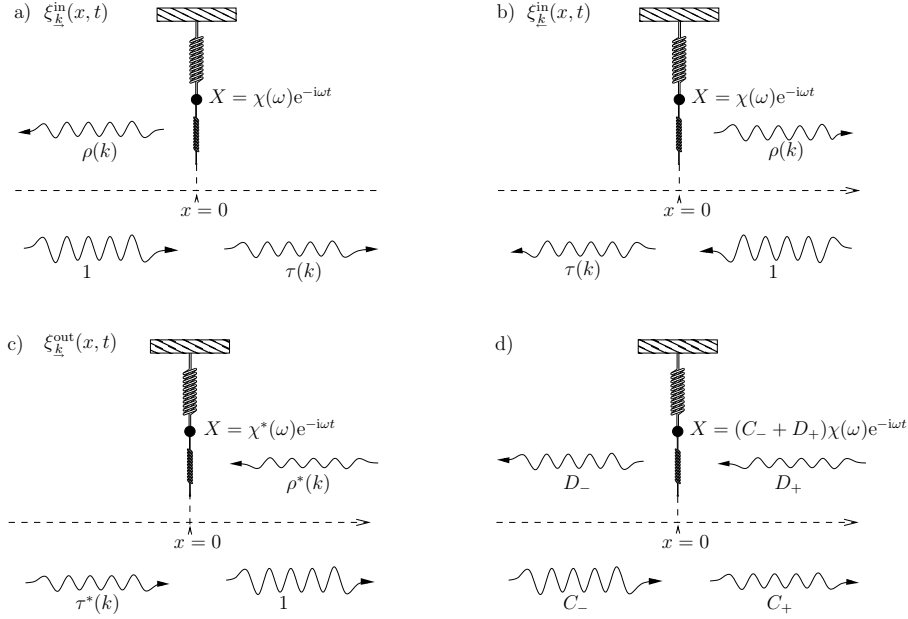}
\caption{\label{fig:inoutT} Several choices for the normal modes with frequency~$\omega$. The wavy arrows symbolise
monochromatic traveling waves whose complex amplitude is specified below. 
a) the in-state given by~\eqref{eq:inmodekp}; 
b) the in-state given by~\eqref{eq:inmodekm};
c) the out-state given by~\eqref{eq:outmodekp} and 
d)~shows the general coefficients 
that will be linked by the $S$ or $T$ matrix defined respectively by~\eqref{def:Smatrix} and~\eqref{def:Tmatrix}.  }
\end{figure}

 When a coupling is present ($\kappa>0$), the two free waves~\eqref{eq:freemode} are not 
 solutions of \eqref{subeq:motiona} anymore. A monochromatic
traveling wave will be partially reflected (resp. transmitted) by the 
oscillator with a reflection (resp. transmission) coefficient~$\rho$ 
(resp.~$\tau$) that is a complex function of~$k$ or $\omega$. The linearity of equations
\eqref{subeq:motiona} guarantees that the frequency will be unchanged by scattering. Indeed, 
to describe such a scattering process, a relevant choice for the two modes of frequency~$\omega$
is to look for in-states, defined (for real positive~$k$) to be of the form (see figure~\ref{fig:inoutT} a)):
\begin{subequations}\label{subeq:inmodek}
\begin{equation}\label{eq:inmodekp}
  \xi^{\mathrm{in}}_\kp(x,t)=\frac{1}{\sqrt{2\pi}}
  \begin{cases}\EXP{\imat(kx-\omega t)}+\rho\, \EXP{\imat(-kx-\omega t)}
               &\mathrm{for}\ x\leqslant0\;;\\
          \tau\, \EXP{\imat(kx-\omega t)}
          &\mathrm{for}\ x\geqslant0\;;
 \end{cases}
\end{equation}
and, since the scattering is symmetric with respect to~$x\mapsto-x$, 
\begin{equation}\label{eq:inmodekm}
  \xi^{\mathrm{in}}_\km(x,t)\DEF
  \xi^{\mathrm{in}}_\kp(-x,t)=\frac{1}{\sqrt{2\pi}}
  \begin{cases}  \tau\, \EXP{\imat(-kx-\omega t)}
               &\mathrm{for}\ x\leqslant0\;;\\
   \EXP{\imat(-kx-\omega t)}+\rho\, \EXP{\imat(kx-\omega t)}     
          &\mathrm{for}\ x\geqslant0\;;
 \end{cases}
\end{equation}
\end{subequations}
with, for both modes, the same amplitude~$\chi(\omega)$ for the oscillator
that can be interpreted, using the language of linear response theory, as a susceptibility:
\begin{equation}\label{eq:X}
  X^{\mathrm{in}}(t)=\chi(\omega)\,\EXP{-\imat\omega t}\;.
\end{equation}
These modes form a set of waves that is suited for
constructing localised wave-packets that
look like free wave-packets when~$t\to-\infty$ (\textit{i.e.} far
away from the oscillator). For instance, when considering a localised wave-packet traveling
to the right, we shall have\footnote{This can be understood with the stationary phase approximation.
If~$\tilde\varphi$ is concentrated around~$k_0>0$, the wave-packets travel with the group 
velocity~$\pm\dmat \omega/\dmat k (k_0)\gtrless0$. When~$t\to-\infty$,  the dominant contributions
to the integrals in \eqref{eq:wavepacket}
form one wave-packet located in~$x<0$ and 
traveling to the right. }: 
\begin{equation}\label{eq:wavepacket}
  \int\tilde\varphi(k)\,\xi^{\mathrm{in}}_\kp(x,t)\,\dmat k
\xrightarrow[t\to-\infty]{}
\int\tilde\varphi(k)\,\xi^{\mathrm{fr}}_k(x,t)\,\dmat k=
\frac{1}{\sqrt{2\pi}}\int\tilde\varphi(k)\,\EXP{\imat(kx-\omega t)}\dmat k\;.
\end{equation}  
The continuity of~$\xi$ at~$x=0$ implies that
\begin{equation}\label{eq:1rhotau}
  1+\rho=\tau\;.
\end{equation}
Conservation of energy implies that the average energetic current is 
conserved in the stationary regime. From \eqref{eq:je}, we must have
\begin{equation}\label{eq:energyconservation}
  1=|\rho|^2+|\tau|^2\;.
\end{equation}
Moreover, the equations are real and invariant under the time reversal, therefore if~$\xi(x,t)$ is a 
solution, so is its complex conjugate~$(\xi(x,t))^*$ and~$\xi(x,-t)$. If we consider the solution 
$\big(\xi^{\mathrm{in}}_\km(x,-t)\big)^*$, then we obtain the mode depicted in~\ref{fig:inoutT} c)
obtained from~\ref{fig:inoutT} b) by reversing the orientation of the arrows and by conjugating the amplitudes. 
This procedure defines
the out-modes that behave like
free modes for remote future times when packed in localised superpositions. We will have
(see figure~\ref{fig:inoutT} c))
\begin{subequations}\label{subeq:outmodek}
\begin{equation}\label{eq:outmodekp}
  \xi^{\mathrm{out}}_\kp(x,t)\DEF\left(\xi^{\mathrm{in}}_\km(x,-t)\right)^*
 =\frac{1}{\sqrt{2\pi}}
  \begin{cases}  \tau^*\, \EXP{\imat(kx-\omega t)}
               &\mathrm{for}\ x\leqslant0\;;\\
   \EXP{\imat(kx-\omega t)}+\rho^*\, \EXP{\imat(-kx-\omega t)}     
          &\mathrm{for}\ x\geqslant0\;,
 \end{cases}
\end{equation}
and
\begin{equation}
  \xi^{\mathrm{out}}_\km(x,t)\DEF\left(\xi^{\mathrm{in}}_\kp(x,-t)\right)^*
 =\frac{1}{\sqrt{2\pi}}
 \begin{cases}\EXP{\imat(-kx-\omega t)}+\rho^*\, \EXP{\imat(kx-\omega t)}
               &\mathrm{for}\ x\leqslant0\;;\\
          \tau^*\, \EXP{\imat(-kx-\omega t)}
          &\mathrm{for}\ x\geqslant0\;;
 \end{cases}
\end{equation}
\end{subequations}
with 
\begin{equation}
  X^{\mathrm{out}}(t)=\chi^*(\omega)\,\EXP{-\imat\omega t}\;.
\end{equation}
If we interpret~$\xi^{\mathrm{out}}_\kp$ as a superposition of in-modes coming from both sides 
that conspire to product
no wave traveling to the right for~$x<0$, we get 
\begin{equation}\label{eq:rhotaustar}
  \tau^*=\phantom{+}\frac{\tau}{\rho+\tau}\;;\qquad
  \rho^*=-\frac{\rho}{\rho+\tau}\;.
\end{equation}

More mathematically, $\xi^{\mathrm{out}}_\kp(x,t)$  can be seen as the
continuation of~$\xi^{\mathrm{in}}_\km(x,t)$  to the domain of negative~$k$'s. Indeed, we have
$\xi^{\mathrm{out}}_\kp(x,t)=\xi^{\mathrm{in}}_\mkm(x,t)$ provided that we 
define
\begin{equation}\label{eq:rhotaumk}
  \tau(-k)\DEF\big(\tau(k)\big)^*\;;\qquad
  \rho(-k)\DEF\big(\rho(k)\big)^*\;.
\end{equation}

\subsection{Definitions of the scattering and transfer matrices}\label{subsec:STdef}

The in-modes and the out-modes are two possible bases
for describing a scattered wave-packet. These bases can be obtained one from each other
by linear transformations; the linearity of the equations of our model 
 implies that they connect waves with the same frequency only, which is a major simplification. 
A typical scattering experiment consists in preparing one wave-packet traveling
 towards the scatterer (the oscillator). Long before the diffusion,
this ingoing wave-packet is a simple superposition of in-modes. Long after the diffusion,
we get two outgoing wave-packets that are naturally described in term of out-modes.
The passage from the in-basis to the out-basis is described in term of the scattering
matrix~$S$ that encapsulates all the information about the possible scattering 
processes\footnote{In the literature, specially within the context of scattering of quantum waves
\cite[\S~2c, for instance]{Taylor72a}, the matrices that connect the free
 waves to the in-waves on the one hand and
the free waves to the out-waves  
on the other hand are often introduced under the name of M{\o}ller operators with the caveat that 
unlike the free states, the set of scattering states may be incomplete, that is insufficient to construct
\textit{all} the states. As we will see in 
\S~\ref{subsec:KGcase}, to get a complete basis one may add to the in-states~\eqref{subeq:inmodek}
the bounded modes when existing. }.
It is made of $2\times2$ blocks~$S(\omega)$ defined by
\begin{equation}
  \begin{pmatrix}\xi^{\mathrm{out}}_\kp\\[1ex]\xi^{\mathrm{out}}_\km
\end{pmatrix}
=S(\omega)
\begin{pmatrix}\xi^{\mathrm{in}}_\kp\\[1ex]\xi^{\mathrm{in}}_\km
\end{pmatrix}.
\end{equation}
The decomposition of each out-mode in term of the two in-modes for, say, $x\leqslant0$, leads to
\begin{equation}\label{eq:Srhotau}
  S(\omega)=
\begin{pmatrix}\dst\frac{\tau}{\rho+\tau}&\dst-\frac{\rho}{\rho+\tau}\\[1em]
               \dst-\frac{\rho}{\rho+\tau}&\dst \frac{\tau}{\rho+\tau}
\end{pmatrix}
=\begin{pmatrix}\tau^* & \rho^*\\ \rho^*&\tau^*
\end{pmatrix}.
\end{equation}
In other words, for the general monochromatic wave 
\begin{equation}\label{def:Smatrix}
  \xi(x,t)=
  \begin{cases}  C_-\EXP{\imat(kx-\omega t)}+D_-\EXP{\imat(-kx-\omega t)}
               &\mathrm{for}\ x\leqslant0\;;\\
              C_+\EXP{\imat(kx-\omega t)}+D_+\EXP{\imat(-kx-\omega t)}    
          &\mathrm{for}\ x\geqslant0\;,
 \end{cases}
\end{equation}
the $S$ matrix connects linearly the coefficients:
\begin{equation}\label{eq:SCD}
   \begin{pmatrix}C_-\\D_+
\end{pmatrix}
=S(\omega)
\begin{pmatrix}C_+\\D_-
\end{pmatrix}.
\end{equation}
 In the absence of scattering ($\tau=1$, $\rho=0$), $S$ simply reduces to the identity.
The unitarity of~$S$, which can be checked on~\eqref{eq:Srhotau},
can be seen as a direct consequence of the conservation of energy since, from~\eqref{eq:je},
the norm of the two vectors involved in~\eqref{eq:SCD} is preserved: $|C_+|^2+|D_-|^2=|C_-|^2+|D_+|^2$.
 
If one wants to calculate the diffusion by several scatterers, it is more convenient 
to introduce the transfer matrix~$T$ whose $2\times2$ blocks are defined
to connect the left coefficients to the right coefficients,
\begin{equation}\label{def:Tmatrix}
  \begin{pmatrix}C_+\\D_+
\end{pmatrix}
=T(\omega)
\begin{pmatrix}C_-\\D_-
\end{pmatrix}.
\end{equation}
Then we have
\begin{equation}\label{eq:Trhotau}
  T(\omega)=
\begin{pmatrix}\dst1+\frac{\rho}{\tau}&\dst\frac{\rho}{\tau}\\[1em]
               \dst-\frac{\rho}{\tau}&\dst \frac{1}{\tau}
\end{pmatrix}
\end{equation}
whose determinant is one.
The addition of one scatterer on the string corresponds to a multiplication by a~$T$ matrix.

\subsection{Physical interpretation of the solutions -- Resonant scattering} 
The definitions and the general properties presented in~\S\S~\ref{subsec:inoutdef} and \ref{subsec:STdef}
are valid for any non-dissipative punctual scatterer. 
As far as our model is concerned, inserting the expression~\eqref{eq:inmodekp} with the 
oscillation~\eqref{eq:X} in equations
\eqref{subeq:motiona} yields to a linear system that can be solved straightforwardly:
\begin{equation}\label{eq:tau}
  \tau(\omega)=\frac{1}{2}+\frac{1}{2}\EXP{-2\imat\eta(\omega)}=\frac{1}{\dst 1
                        +\imat\frac{\kappa}{2\sqrt{\omega^2-\omega_0^2}}\,
                         \frac{\omega^2-\Omega_0^2}{\omega^2-\Omega_\kappa^2}}\;;
\end{equation}
\begin{equation}\label{eq:rho}
  \rho(\omega)=-\frac{1}{2}+\frac{1}{2}\EXP{-2\imat\eta(\omega)}
              =\frac{-1}{\dst 1
                         -\imat\frac{2\sqrt{\omega^2-\omega_0^2}}{\kappa}\,
                          \frac{\omega^2-\Omega_\kappa^2}{\omega^2-\Omega_0^2}}\;,
\end{equation}
with
\begin{equation}\label{eq:theta}
  \eta(\omega)\DEF\arctan\left(\frac{\kappa}{2\sqrt{\omega^2-\omega_0^2}}\,
                         \frac{\omega^2-\Omega_0^2}{\omega^2-\Omega_\kappa^2}\right)
\end{equation}
and
\begin{equation}\label{eq:chi}
  \chi(\omega)=\frac{\kappa/\sqrt{2\pi}}{\dst\Omega_\kappa^2-\omega^2
                    -\imat\frac{\kappa}{2\sqrt{\omega^2-\omega_0^2}}(\omega^2-\Omega_0^2)}\;.
\end{equation}

Even though the coefficients~$\rho$ and~$\tau$ were first defined for traveling waves, \textit{i.e.}
for~$\omega>\omega_0$, the above expressions can be continued for~$\omega<\omega_0$. We will
understand the physical interpretation of this procedure when we will study radiation in~\S~\ref{sec:radiation}.
As long as~$\omega>\omega_0$, \eqref{eq:energyconservation}, \eqref{eq:rhotaustar}
and \eqref{eq:rhotaumk} hold.
We can check that the ultra-violet limit~$\omega\to\infty$ is equivalent to the limit of weak 
coupling where the oscillator becomes transparent:~$\tau\to1$ and~$\rho\to0$. Another case where
the coupling is inefficient is when~$A$ and~$X$ both oscillate in phase with~$\omega=\Omega_0$ since the
 coupling spring remains unstretched. 
\begin{figure}[!ht]
\center
\includegraphics[width=\textwidth]{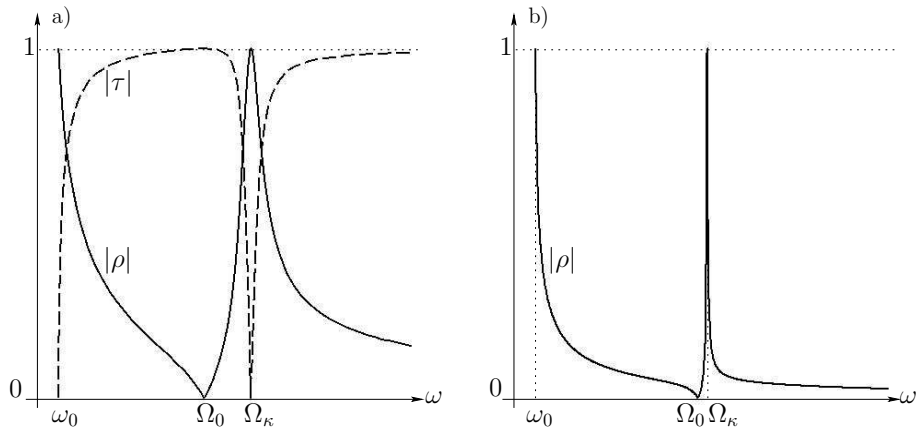}
\caption{\label{fig:rhotau} Graphs of $|\tau(\omega)|$ (dashed line) 
and~$|\rho(\omega)|$ (solid line) given by~\eqref{eq:tau} and~\eqref{eq:rho}
when~$\omega_0<\Omega_0$. One can observe a resonance scattering
 spike for~$\omega=\Omega_k$ and an antiresonance for~$\omega=\Omega_\kappa$ where no scattering occurs.
The quality factor is a) $Q\simeq15$ and b)~$Q\simeq200$. Let us mention 
that when~$\Omega_0<\omega_0<\Omega_\kappa$, the antiresonance has vanished
and the local maximum~$|\rho(\Omega_k)|=1$ is too soft to be called a resonance. When
$\Omega_\kappa<\omega_0$, no scattering resonance
occurs and~$|\rho|$ decreases monotonically from 1 at~$\omega=\omega_0$ to 0 when~$\omega\to\
+\infty$. }
\end{figure} 
More interesting is the resonant scattering that occurs, provided that $\omega_0<\Omega_\kappa$,
 when the ingoing wave that forces the oscillator
 has precisely the
same frequency as the shifted frequency of the latter, that is~$\omega=\Omega_k$. Then, the scattering is
the most efficient since no transmission occur ($\rho=-1$, $\tau=0$). The resonance spike can be 
seen in figure~\ref{fig:rhotau} and its quality factor can be evaluated from its width~$\Delta\omega$
when~$|\rho(\Omega_\kappa\pm\Delta\omega/2)|=1/\sqrt{2}$: for a small coupling,
\begin{equation}
  Q\DEF\frac{\Omega_\kappa}{\Delta\omega}=\frac{2\Omega_0^2}{\kappa^2}\sqrt{\Omega_0^2-\omega_0^2}\,\big(1+\Omat(\kappa)\big)\;,
\end{equation}  
 and therefore, the smaller~$\kappa$, the better the quality of the resonance.

\section{Radiation, damping and bounded mode}\label{sec:radiation}

The general idea that damping and therefore irreversibility emerge because of
the interaction with a large number of degrees of freedom can be illustrated explicitly on our model.
If we choose initial conditions such that the spring is at rest at $t=0$, the entire energy being
contained in the oscillator, for instance 
\begin{equation}\label{eq:ic}
  \xi(x,0)=0\;;\quad  \frac{\partial\xi}{\partial t}(x,0)=0\;;\quad  X(0)=X_0\;;\quad  \dot{X}(0)=0\;,
\end{equation}
the energy transfer to the string will damp the oscillations of~$M$ and the latter may 
completely lose its energy far before the energy can get back from the string if its boundary is far
away from the oscillator (for a string of length~$\ell$, Poincar\'e recurrence time is of 
order $\ell/c$ if there were no dispersion). The equation of motion of the oscillator
is particularly simple when the string is non-dispersive ($\omega_0=0$) and therefore 
we will start by studying this case. However, we will also
consider the case of the Klein-Gordon string because  when~$\omega_0>\Omega_0$,
 we will see that there exists a stable mode of the oscillator at 
a frequency~$\omega_b>0$ whose dissipation is blocked because~$\omega_b<\omega_0$.
Its vibration does not decay because at this frequency, only evanescent waves 
can exist on the string, which do not carry away energy current on average~(see~\eqref{eq:je_ev}).

\subsection{The d'Alembert string}

With the initial conditions~\eqref{eq:ic}, the general form of the radiated waves on the string
will be~$\xi(x,t)=\xi_0(t-|x|)$: each of the two wave-packets travels
 away from~$x=0$ without distortion when~$\omega_0=0$. Equation~\eqref{eq:discontinuityslopea} becomes
\begin{equation}\label{eq:xi0X}
  2\dot{\xi}_0+\kappa\xi_0=\kappa X
\end{equation}
and the elimination of~$\xi_0$ from~\eqref{eq:xi0X} and \eqref{eq:motionMa} yields to
\begin{equation}\label{eq:AbrahamLorentz}
  \ddot{X}+\Omega_0^2X=-\frac{2\Omega_\kappa^2}{\kappa}\dot{X}-\frac{2}{\kappa}\dddot{X}\;.
\end{equation}
The two terms in the right hand side are dissipative forces. The first
one has the familiar taste of the viscous resistive force whereas the
second has the flavour of the Schott term~$2e^2\,\dddot{x}/3$
\cite[eq. (2.7b)]{Rohrlich00a} in the Abraham-Lorentz-Dirac equation
which governs the dynamics of an electric charge~$e$ that takes into
account the electromagnetic self-force of the charge. The major difference is
the sign of the coefficient in front of the third-derivative. Unlike the Schott term,
the negative sign in~\eqref{eq:AbrahamLorentz} prevents the spurious exponentially accelerating solutions.
It can be clearly seen that the irreversibility due to dissipation comes straightforwardly from the choice
of initial conditions \eqref{eq:ic} that break the time-reversal symmetry under which the original equations
are invariant. 

Many models of an oscillator coupled to one-dimensional waves are
recovered in the limit of strong coupling~$\kappa\to+\infty$ (see the
references given in the introduction, for instance when the mass is
directly attached on the string). In that case, only the viscous
force remains in~\eqref{eq:AbrahamLorentz} and we immediately get the
well-known damped oscillator $\ddot{X}+2\dot{X}+\Omega_0^2X=0$.

Looking for exponential solutions~$X(t)=X(0)\EXP{zt}$ yields to the characteristic equation of~\eqref{eq:AbrahamLorentz}:
\begin{equation}
  z^3+\frac{1}{2}\kappa z^2+\Omega_\kappa^2 z+ \frac{1}{2}\kappa\Omega_0^2=0\;.
\end{equation}
There are three solutions, one real $z_0$ and two complex~$z_+, z_-$ all having a strictly 
negative real part.
Perturbatively in~$\kappa$, we have
\begin{subequations}
\begin{eqnarray}
  z_0&=&-\frac{\kappa}{2}+\frac{\kappa^2}{2\Omega_0^2}+\Omat(\kappa^3)\;;\\
  z_+&=&z_-^*\ =\ -\frac{\kappa^2}{4\Omega_0^2}+\imat\left(\Omega_0+\frac{\kappa}{2\Omega_0}-\frac{\kappa^2}{8\Omega_0^3}\right)
 + \Omat(\kappa^3)\;.
\end{eqnarray}
\end{subequations}
For generic initial conditions, including~\eqref{eq:ic}, where the string is at rest the energy of the oscillator
will exponentially decay like~$\EXP{-\Gamma t}$ at the rate
\begin{equation}\label{eq:Gamma}
  \Gamma=\frac{\Omega_0}{Q}=\frac{\kappa^2}{2\Omega_0^2}+\Omat(\kappa^3)\;.
\end{equation}
In the language of particle physics, the stable non-interacting particle (the mode of the free oscillator)
has been destabilised into a metastable particle of lifetime~$\Gamma^{-1}$ because of its interactions.

\subsection{The Klein-Gordon string}\label{subsec:KGcase}

When~$\omega_0>0$, one cannot get a differential equation for~$X(t)$ but must keep working with its 
temporal Fourier transform
\begin{equation}\label{eq:Xomega}
  \tilde{X}(\omega)\DEF\frac{1}{\sqrt{2\pi}}\int X(t)\,\EXP{\imat\omega t}\dmat t\;,
\end{equation}
together with a superposition of purely radiated waves of the 
form~$\xi(x,t)=(\sqrt{2\pi})^{-1}\int\tilde{\xi}(\omega)\,\EXP{\imat(k|x|-\omega t)}\dmat\omega$.
 Inserting them in \eqref{subeq:motiona},
$\tilde{X}$ must satisfy\footnote{The presence of the square root in~\eqref{eq:cubicomega} is the
reason  that prevents us from
obtaining a local differential operator for~$X(t)$.}
\begin{equation}\label{eq:cubicomega}
  \left[\kappa(\omega^2-\Omega_0^2)-2\imat\sqrt{\omega^2-\omega_0^2}(\omega^2-\Omega_\kappa^2)\right]\tilde{X}(\omega)=0\;.
\end{equation}
Therefore~$\tilde{X}$ vanishes everywhere but at the frequencies that
cancel the brackets. These are precisely the poles of~$\tau$
and therefore of~$\rho=\tau-1$ given by~\eqref{eq:tau} and~\eqref{eq:rho}. Indeed, for
pure radiative modes, the ingoing waves vanish ($C_-=D_+=0$) and
therefore the matrix element~$T_{22}$ must go to infinity in order to
keep~$D_-$ finite (see figure~\ref{fig:inoutT} d) and
equations~\eqref{def:Tmatrix}
and~\eqref{eq:Trhotau}). Letting~$Z=-\omega^2$, we look for the solutions
of the cubic equation
\begin{equation}\label{eq:cubicZ}
  (Z+\omega_0^2)(Z+\Omega_0^2+\kappa)^2-\frac{\kappa^2}{4}(Z+\Omega_0^2)^2=0\;.
\end{equation}
Perturbatively in~$\kappa$, those are
\begin{subequations}
\begin{eqnarray}
  Z_0&=&-\omega_0^2+\frac{\kappa^2}{4}-\frac{\kappa^3}{2(\Omega_0^2-\omega_0^2)}+\Omat(\kappa^4)\;;\\
  Z_+&=&Z_-^*= -\Omega_0^2-\kappa-\imat\frac{\kappa^2}{2\sqrt{\Omega_0^2-\omega_0^2}}+ \Omat(\kappa^3)\;.
\end{eqnarray}
\end{subequations}
When~$\omega_0\to0$, we recover~$Z_0\to z_0^2$ and~$Z_\pm\to
z^2_\pm$. The physical frequencies will be the three square
roots~$\imat\sqrt{Z_0}$,$\imat\sqrt{Z_\pm}$ whose imaginary part is
not positive: The typical decay rate of energy will be given by the
nearest root~$\omega_\mathrm{min}$ to the real axis:
$\Gamma=-2\mathrm{Im}(\omega_\mathrm{min})$.  As long
as~$2\omega_0<\kappa\ll\Omega_0^2$, all the three frequencies have
strictly negative real part. The decay rate is given by
\begin{equation}
  \Gamma=\frac{\Omega_0}{Q}=\frac{\kappa^2}{2\Omega_0\sqrt{\Omega_0^2-\omega^2_0}}+\Omat(\kappa^3)\;.
\end{equation}
It is a very general feature that the poles of the~$S$ matrix are
associated with resonances and, more precisely, that their imaginary
part provide the decay rates, which are proportional to the inverse of
the quality factor of the resonances.

When~$\kappa<2\omega_0$, $Z_0$ is negative, one residual
oscillation persists at frequency~$\omega_b\DEFt\sqrt{-Z_0}$.
  For~$\omega_b<\omega_0$, no transfer of energy is
allowed; only evanescent waves are created and those do not carry any
average energy current. Unlike the scattering states, this
non-decaying mode is spatially localised. More generally, any bounded
mode has a purely real frequency~$\omega_b$ that must be less
than~$\omega_0$ since~$Z_b+\omega_0^2=\omega_0^2-\omega_b^2>0$ in
order to fulfil~\eqref{eq:cubicZ}. From~\eqref{eq:dispersion},
$k(\omega_b)$ is therefore purely imaginary. Moreover, in order to
cancel the bracket in~\eqref{eq:cubicomega}, $\omega_b^2$ must lie in
between~$\Omega^2_0$ and~$\Omega^2_\kappa$. For simplicity, let us
introduce the auxiliary parameter
$\Upsilon\DEFt(\Omega_0^2-\omega_0^2)/\kappa$ and the real positive
variable~$u\DEFt|k|/\sqrt{\kappa}$; a stable mode will exist if the
cubic equation
\begin{equation}\label{eq:cubicu}
  2u(u^2+\Upsilon+1)+\sqrt{\kappa}(u^2+\Upsilon)=0\;.
\end{equation}
has a positive real solution. This can be achieved for~$\Upsilon<0$
only \textit{i.e.}  in a regime where~$\Omega_0<\omega_0$. For
$\Upsilon<0$, the product of the roots of the left hand side
of~\eqref{eq:cubicu}, $u_+u_-u_b$ is~$-\sqrt{\kappa}\Upsilon>0$.  If
two roots are complex conjugated, therefore the third one is
necessarily positive. If the three roots are real, either only one is
positive or all three of them are. The latter case must be ruled
out since the sum~$u_++u_-+u_b=-\sqrt{\kappa}$ is strictly negative.  We have therefore
proved that a sufficient and necessary condition for a stable mode to
exist is that~$\omega_0>\Omega_0$. Its frequency is given
by~
\begin{equation}\label{eq:omegab}
  \omega_b=\sqrt{\omega_0^2-\kappa u^2_b}
\end{equation}
 where $u_b$ is the unique
positive real solution of~\eqref{eq:cubicu}. 
Perturbatively in~$\kappa$, we have
\begin{equation}
  \omega_b=\Omega_0+\frac{\kappa}{2\Omega_0}
  -\frac{2\Omega_0^2+\sqrt{\omega_0^2-\Omega_0^2}}{\Omega_0^3\sqrt{\omega_0^2-\Omega_0^2}}\;\frac{\kappa^2}{8}+\Omat(\kappa^3)
\end{equation}
and the corresponding bounded mode is given by
\begin{subequations}\label{subeq:bmode}
\begin{equation}
  \xi^{b}(x,t)=C_b\,\EXP{-\sqrt{\omega_0^2-\omega_b^2}|x|}\EXP{-\imat\omega_b t}\; ;
\end{equation}
\begin{equation}
  X_b(t)=\frac{\kappa\, C_b}{\Omega^2_\kappa-\omega_b^2}\,\EXP{-\imat\omega_b t}\;.
\end{equation}
\end{subequations}
The choice of the normalization,
\begin{equation}\label{eq:A}
  C_b=\left(\frac{1}{\sqrt{\omega_0^2-\omega_b^2}}+\frac{\kappa^2}{\dst(\Omega_\kappa^2-\omega_b^2)^2}\right)^{-1/2}
\end{equation}
 will be justified below (equation \ref{eq:Xibnorm}).

\section{Some like it diagonal}\label{sec:hamdiag}

\subsection{Normal coordinates}
 
What makes the model completely tractable is of course that it remains
linear. However, the direct diagonalization of the quadratic
Hamiltonian~\eqref{def:Ham} remains particularly difficult.  In that
case, the trick is to solve the equations of motion to determine the
normal modes first --- this is precisely what we have done in the previous
paragraphs --- and then write the Hamiltonian in its diagonal form,
\begin{equation}\label{eq:Haa}
 H=\frac{1}{2}\sum_{\alpha}\left(p_\alpha^2+\omega_\alpha^2q_\alpha^2\right)
  =\sum_{\alpha}\omega_\alpha\; a_\alpha^*\,a^{}_\alpha
\end{equation}
in term of some (real) canonical coordinate~$\{p_\alpha,q_\alpha\}_\alpha$ or 
(complex) normal coordinates~$\{a_\alpha\}_\alpha$ that are 
 associated with modes 
labelled by the discrete and/or continuous index~$\alpha$\footnote{To avoid ambiguities we will often 
subscript the brace describing a set like~$\{\dots\}_{\alpha\in \mathcal{A}}$ to recall
which indices are running  and what is their range~$\mathcal{A}$ if the latter does matter.}. We have
\begin{equation}
  a_\alpha=\sqrt{\frac{\omega_\alpha}{2}}\,q_\alpha +\frac{\imat}{\sqrt{2\omega_\alpha}}\,p_\alpha\;;
\end{equation}
\begin{equation}
  q_\alpha=\frac{1}{\sqrt{2\omega_{\alpha}}}(a^*_\alpha+a^{}_\alpha)\;;\qquad
  p_\alpha=\imat\sqrt{\frac{\omega_\alpha}{2}}(a^*_\alpha-a^{}_\alpha)\;;
\end{equation}
and, for each pair~$\{\alpha_1,\alpha_2\}$,
\begin{equation}
  \Poisson{a_{\alpha_1}}{a_{\alpha_2}}\ =\ 0\;;\qquad
  \Poisson{a^{}_{\alpha_1}}{a_{\alpha_2}^*}\ =\ \imat\delta_{\alpha_1,\alpha_2}\;;
\end{equation}
\begin{equation}
  \Poisson{p_{\alpha_1}}{p_{\alpha_2}}\ =\ 0\;;\quad
  \Poisson{q_{\alpha_1}}{q_{\alpha_2}}\ =\ 0\;;\quad
  \Poisson{p_{\alpha_1}}{q_{\alpha_2}}\ =\ \delta_{\alpha_1,\alpha_2}\;;
\end{equation}
where~$\delta$ stands for the Kronecker symbol or the Dirac distribution. 
The second step consists in determining the canonical transformation
that expresses $a_\alpha$ in terms of some a priori known 
normal coordinates, namely some free normal modes~$a_{\mathrm{fr},\alpha}$. 
In our case this transformation is linear and will be transposed directly into
the quantum theory by replacing the complex number~$a_\alpha$ (resp. $a^*_\alpha$) by the 
creation (resp. annihilation) operator~$\hat{a}_\alpha$ (resp. its Hermitian conjugate~$\hat{a}^*_\alpha$)
 of the~$\alpha$th one-particle 
eigenstate whose energy is~$\hbar\omega_\alpha$. 
As we have seen, all the scattering states are  twice degenerate, in the sense 
that each normal frequency~$\omega$ is associated with two independent states labelled
by~$k$ and~$-k$. These modes both diagonalize the Hamiltonian~\eqref{def:Ham}.
An infinite number of pairs of eigenvectors 
can be chosen to constitute a basis, among them, the in and out-states,
which are particularly relevant as soon as we get into a quantum field theory\footnote{\label{fn:qft}
The normal coordinates~$a_{\mathrm{in}}(k)$ and~$a_{\mathrm{out}}(k)$ 
 constructed from the scattering modes, once quantized into~$\hat{a}_{\mathrm{in}}(k)$ 
and~$\hat{a}_{\mathrm{out}}(k)$, allow the interpretation  of the quantum states
in term of asymptotic (quasi-)particles; more precisely the linear transformations 
from the free~$\hat{a}_{\mathrm{fr}}(k)$
provide the explicit connection between the non-interacting states (the Fock space
for bare particles including the free vacuum)
and the interacting states (the Fock space for dressed particles including 
the interacting vacuum).}.
But it order to get~\eqref{eq:Haa} properly one must check that the set of modes
is actually complete --- \textit{i.e.} that any kind of motion of our system
can be described as a linear superposition of modes --- and orthonormalized correctly in order to deal with canonical
complex coordinates. 
 Fourier analysis assures that the free states~\eqref{eq:freemode} constitute a complete set for describing
the waves on the string. When interacting with the oscillator, 
if~$\omega_0>\Omega_0$ one bounded state exists that must be added to the in-modes (or to the out-modes) to
get a genuine basis. Then, including the normal coordinates~$A_b$ of the bounded mode if there is any,
\eqref{eq:Haa} reads
\begin{equation}
 H=\omega_b\,A_b^*A^{}_b+\!\!\int\!\omega(k)\;a_{\mathrm{in}}^*(k)\,a^{}_{\mathrm{in}}(k)\,\dmat k
=\omega_b\,A_b^*A^{}_b+\!\!\int\!\omega(k)\;a_{\mathrm{out}}^*(k)\,a^{}_{\mathrm{out}}(k)\,\dmat k\;.
\end{equation}
We chose the convention that, when~$k>0$, $a_{\mathrm{in}}(k)$ (resp. $a_{\mathrm{out}}(k)$) is constructed from
$\xi^{\mathrm{in}}_\kp$  (resp.~$\xi^{\mathrm{out}}_\kp$) while $a_{\mathrm{in}}(-k)$ (resp. $a_{\mathrm{out}}(-k)$)
 is constructed from
$\xi^{\mathrm{in}}_\km$  (resp.~$\xi^{\mathrm{out}}_\km$).
\subsection{Orthonormalization}

Having a complete set of modes does not guarantee that they are orthogonal. Indeed, it may happen
that two eigenvectors having a common eigenfrequency are not. For instance, one must check in one way or another
that the modes~\eqref{subeq:inmodek} are orthogonal and properly normalized. If we denote by~$\Xi(t)$ a \textit{classical}
state represented by  the displacement~$X(t)$ of the oscillator and the wave~$\xi(x,t)$ on the string, 
the scalar product between two states~$\Xi_1(t)$ and~$\Xi_2(t)$ is 
\begin{equation}
  \Xi_1(t)\cdot\,\Xi_2(t)\DEF X_1^*(t)X^{}_2(t)+\int\xi_1^*(x,t)\,\xi^{}_2(x,t)\,\dmat x\;.
\end{equation}
It is shown in the appendix that if  $\Xi_k^{\mathrm{in}}$ (resp.~$\Xi_{-k}^{\mathrm{in}}$ ) stands for 
the mode $\xi^{\mathrm{in}}_\kp(x,t)$ (resp.~$\xi^{\mathrm{in}}_\km(x,t)$)
 both with~$X^{\mathrm{in}}(t)=\chi(\omega)\EXP{-\imat\omega t}$, then we have, for any (positive and/or negative) real 
pair~$(k_1,k_2)$,
\begin{equation}
  \Xi_{k_1}^\mathrm{in}(t)\cdot\,\Xi_{k_2}^\mathrm{in}(t)=\delta(k_1-k_2)\;.
\end{equation}
It is easy to see that we chose the normalization \eqref{eq:A} in order to get 
\begin{equation}\label{eq:Xibnorm}
  \Xi^b(t)\cdot\,\Xi^b(t)=1\;.
\end{equation}
The free states $\Xi_{\pm k}^{\mathrm{fr}}$ represented by~$X\mathop{=}0$ and~\eqref{eq:freemode} are clearly 
orthonormalized,~$\Xi_{k_1}^{\mathrm{fr}}\cdot\,\Xi_{k_2}^{\mathrm{fr}}=\delta(k_1-k_2)$, and form a complete basis
if we add the state that allow to describe the motion of the oscillator, namely
$\Xi_\mathrm{osc}^\mathrm{fr}$ represented by~$X=1$ and~$\xi\equiv0$. 

\subsection{The real symmetric modes}\label{subsec:realsymmetric}

The potential in~\eqref{def:Ham} is a real definite positive symmetric
quadratic form and therefore can be diagonalized in an orthogonal
basis of \textit{real} vectors. The natural choice of retaining the
real or the imaginary part of~$\Xi_{\pm k}^{\mathrm{in}}$ actually
provides two real modes but that are not orthogonal.  A way to assure
that we deal with an orthogonal basis, is to pick up a symmetry, say
the parity~$x\mapsto-x$, of the Hamiltonian and classify the
eigenmodes accordingly. The bounded state, if there is any, remains even.
  The in and out modes are not symmetric under
space inversion but it is straightforward to obtain eigenmodes that
are also eigenvectors of parity.  For any complex factors~$c_\pm$, the combinations 
$c_\pm(\Xi_{k}^{\mathrm{in}}\pm\Xi_{-k}^{\mathrm{in}})$ are
symmetric/antisymmetric eigenvectors at any time with the
eigenvalues~$\pm1$. After some algebraic manipulations 
using the expressions~\eqref{eq:rho} of~$\rho$ in term of~$\eta$ given by~\eqref{eq:theta}, the
symmetric and antisymmetric modes  are represented, for~$k>0$,
by
\begin{subequations}
\begin{eqnarray}
  c_+\big(\xi_\kp^{\mathrm{in}}(x,t)+\xi_\km^{\mathrm{in}}(x,t)\big)&=&
c_+\,\EXP{-\imat\eta}\,\sqrt{\frac{2}{\pi}}\cos(k|x|-\eta)\,\EXP{-\imat\omega t}\;;
\\
  c_-\big(\xi_\kp^{\mathrm{in}}(x,t)-\xi_\km^{\mathrm{in}}(x,t)\big)&=&
\imat c_-\sqrt{\frac{2}{\pi}}\,\sin(kx)\,\EXP{-\imat\omega t}\;.
\end{eqnarray}
\end{subequations}
The choice~$c_+=\EXP{\imat\eta}/\sqrt{2}$ and~$c_-=-\imat/\sqrt{2}$ leads to the real 
normalized symmetric modes, defined for~$k>0$ by
\begin{equation}\label{eq:XiRXi}
  \begin{pmatrix}\Xi^+_{k}\\ \Xi^-_{k}
\end{pmatrix}
=R(\omega)
\begin{pmatrix}\Xi_{k}^{\mathrm{in}}\\\Xi_{-k}^{\mathrm{in}}
\end{pmatrix}.
\end{equation}
with the unitary matrix
\begin{equation}\label{eq:R}
  R(\omega)=\frac{1}{\sqrt{2}}
\begin{pmatrix}\dst\EXP{\imat\eta(\omega)}&\dst\EXP{\imat\eta(\omega)}\\[1em]
               \dst-\imat&\dst \imat
\end{pmatrix}.
\end{equation}

Constructing the real (anti)symmetric states from the out-modes at any time leads
to the same $\Xi^\pm_{k}$  since the common eigenspace to $H$ and to the
parity is of dimension one. To sum up, for~$k>0$, $\Xi^\pm_{k}$ is represented by
\begin{subequations}
\begin{flalign}\label{eq:xi+}
  \xi^+_{k}(x,t)&=\frac{1}{\sqrt{\pi}}\cos[k|x|-\eta(\omega)]\,\EXP{-\imat\omega t}\;; 
&X^+(t)&=\frac{\kappa}{\sqrt{\pi}}\frac{\cos[\eta(\omega)]}{\Omega_\kappa^2-\omega^2}\,\EXP{-\imat\omega t}\;;\\
  \xi^-_{k}(x,t)&=\frac{1}{\sqrt{\pi}}\,\sin(kx)\,\EXP{-\imat\omega t}\;; &X^-(t)&=0\;,
\end{flalign}
\end{subequations}
and~$\{\Xi_b(t)\}\cup\{\Xi^\pm_{k}(t)\}_{k>0}$ is, at any time, an orthonormalized eigenbasis of symmetric or antisymmetric eigenvectors
of the Hamiltonian with eigenvalues given respectively by \eqref{eq:omegab} and~\eqref{eq:dispersion}. 

\subsection{An atom in a closed cavity}

The explicit canonical linear transformation that connects the free canonical variables to the 
interacting ones (the in or out modes via the real symmetric ones) is beyond the scope of this article
and will be given and extensively interpreted in a future paper where we will quantize our model.
 As explained above (see \S~\ref{subec:hamiltonian} and
 also the note \ref{fn:qft}), this is really
interesting and beyond a purely academic exercise only if one wants to switch to quantum theory and/or
statistical physics. The quantum linear transformation between~$\hat{a}_{\mathrm{in}}(k)$ and~$\hat{a}_{\mathrm{fr}}(k)$
 appears to be a generalised 
Bogoliubov transformation and our model provides an explicit construction of quasi-particles in terms of free particles.  
 \begin{figure}[!ht]
\center
\includegraphics[width=\textwidth]{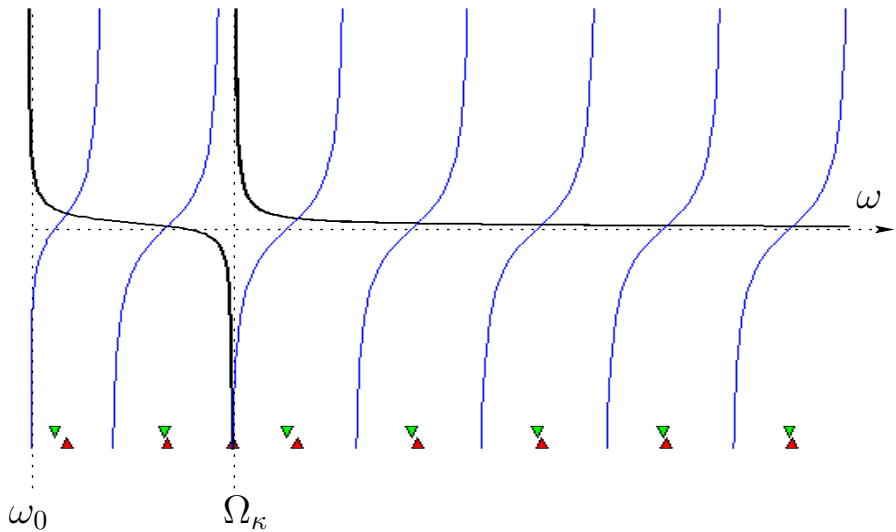}
\caption{\label{fig:cavite}Graphical resolution of equations~\eqref{eq:kncavite} that provide
the even frequency spectrum  when the oscillator is 
attached to the middle of string of finite size~$\ell$. The frequencies, represented by small up triangles  on an  
horizontal line at the bottom of the figure,
are centered on the abscissae of the intersections of the graph $\omega\mapsto\tan\big(\eta(\omega)\big)$ (thick solid line) with
the graphs $\omega\mapsto\tan\big(k(\omega)\ell-\pi/2\big)$ (thin solid lines) for~$\omega>\omega_0$. As
 in figure~\ref{fig:rhotau} a), the quality factor of the resonance is~$Q\simeq15$.
 The down triangles indicate
the spectrum of vibration of the Klein-Gordon string alone of finite size~$\ell$ obtained for 
$\tan\big(k(\omega)\ell/2-\pi/2\big)=0$, that is for~$\omega_n=\sqrt{\omega_0^2+\pi^2(2n+1)^2/\ell^2}$ with~$n$ 
a positive integer.}
\end{figure} 

However, the real symmetric modes that have been founded in the previous section remain interesting at the less advanced
level of the present article because they are the natural modes to work with when 
the finite size~$\ell$ of the string becomes relevant. Indeed, when $\ell/c$ is not too large  
compared to the typical time $\Gamma^{-1}$ characterising the radiations of the oscillator,
the discrete character of the spectrum of the non-interacting string can be ``feeled'' by the oscillator.
When  boundary conditions are imposed, say~$\xi(\ell/2,t)=\xi(-\ell/2,t)=0$,
the discrete (even) spectrum of the whole system is modified by the presence of the oscillator and, from \eqref{eq:xi+}
given by the~$\{k_n\}_{n\in\ZZ}$ that fulfill the equations
\begin{equation}\label{eq:kncavite}
  \frac{1}{2}\,k_n\ell-\eta\big(\omega(k_n)\big)
  =\frac{\pi}{2}+n\pi\quad\Longleftrightarrow\quad \tan\big(\eta(k)\big)=\tan(k\ell/2-\pi/2)
\end{equation}
that can be solved graphically (figure~\ref{fig:cavite}). The frequency~$\Omega_0$ of the
free oscillations of the mass 
inserts in the spectrum of the string. The even
spectrum will differ from the non-interacting case
when~$\EXP{\imat\eta}$ is significantly different from one. For
resonances with high quality, it will not affect the frequencies that
are away from the resonant frequency.

 What one gets here, for~$\omega_0=0$
 is an elementary model of an atom in a (perfect) electrodynamics 
cavity of size~$\ell$ (some imperfections can be taken into account
if we relax the Dirichlet boundary conditions and put partially reflectives ``mirrors'' on the string). 
The field may or may not be quantized and, not 
to speak of lasers, we obtain a sort of primer for the widespread
physics of quantum electrodynamics cavities that 
have been realised 
in laboratory to test successfully some fundamental concepts
in quantum physics \cite{Haroche/Raymond06a}. The purely mechanical model
 for infinite~$\kappa$ (the mass is directly attached on the string) has been carefully
studied with experiments in \cite{Gomez+07a}.

\section{Conclusion}\label{sec:conclusion}

In addition to a more detailed study of the finite size effects, another natural development
of the present work would be to deal with multiple scatterers. For instance, when there are two identical scatterers
with~$\omega_0>\Omega_0$, we expect that the degeneracies of the two bounded modes is broken 
and that the
splitting between the symmetric and the antisymmetric 
bounded modes decreases exponentially with the separation of the oscillators.
Starting with initial conditions where only one oscillator has some energy, the beating between the two oscillators
is an example of tunnelling due to the presence of evanescent waves connecting the two oscillators.

Even before we quantize the whole system, our model may be interesting 
 to keep the field classical whereas only the oscillator is quantized. It would provide an
 illustration of say, the Fermi golden rule within the context of time-dependent 
perturbation theory \cite[chap.~XIII]{CohenTannoudji+80a}. However, we have proven that
this golden rule transpires in our classical model since the 
transition rate~\eqref{eq:Gamma} to the continuum of the modes is proportionnal to the square 
of the coupling strength~$\kappa$ to first order in the perturbation.

As it has been demonstrated, this model captures many fondamental phenomena
that are important in many areas of physics and offers wide possibilities for pedagogical use.
 Above all, I hope it will help the reader to discover and/or to understand them more 
deeply.

\subsection*{Acknowledgments}

Many thanks to Domique Delande and Beno\^{\i}t Gr\'emaud for their hospitality at the Laboratoire
 Kastler-Brossel (Paris) and to Stam Nicolis (Laboratoire de Math\'ematiques et de Physique 
Th\'eorique, Tours) for his careful reading of the manuscript. 

\section{Appendix: Normalization of the modes}
The construction of the real symmetric modes presented in \S~\ref{subsec:realsymmetric}
leads to an orthogonal basis~$\{\Xi_b(t)\}\cup\{\Xi^\pm_{k}(t)\}_{k>0}$. 
When it exists ($\omega_0>\Omega_0$), the bounded state $\Xi_b$ has a norm unity.
 That they are eigenvectors for different eigenvalues of~$H$ or
of the parity guarantees that~$\Xi_{k_1}^\pm\cdot\,\Xi^\pm_{k_2}\propto\delta(k_1-k_2)$
and this appendix proves that the proportionality factor is indeed unity. 

Rewriting~\eqref{eq:inmodekp} with the help of~\eqref{eq:1rhotau} and with the Heaviside step function~$\Theta$,
\begin{equation}
  \xi^{\mathrm{in}}_\kp(x,t)=\frac{1}{\sqrt{2\pi}}
\left[\EXP{\imat kx} + \rho(k)\,\EXP{-\imat kx}\Theta(-x)+ \rho(k)\,\EXP{\imat kx}\Theta(x)\right]\EXP{-\imat\omega t}\;,
\end{equation}
we have, with~$\rho_1\DEF\rho(k_1)$ and $\rho_2\DEF\rho(k_2)$,
\begin{multline}\label{eq:xi1xi2}
 \!\!\! \int\xi_\kpu^*(x,t)\,\xi^{}_\kpd(x,t)\,\dmat x=\delta(k_1-k_2)
+\frac{\imat}{2\pi}\left[
 \rho^{}_2\left(\frac{1}{k_2+k_1+\imat0^+}+\frac{1}{k_2-k_1+\imat0^+}\right)\right.\\
+\rho^*_1\left(\frac{1}{k_2-k_1+\imat0^+}-\frac{1}{k_2+k_1-\imat0^+}\right)
\left.+\frac{2\rho^*_1\rho^{}_2}{k_2-k_1+\imat0^+}
                 \right].
\end{multline}
We have used the identity ($0^+$ stands for the limit~$\epsilon\to0$ keeping~$\epsilon>0$)
\begin{equation}
  \int_{x_0}^{+\infty}\EXP{\imat kx}\,\dmat x=\frac{\imat\EXP{\imat{kx_0}}}{k+\imat0^+}
\end{equation}
valid for any real~$k$. The other identity
\begin{equation}
  \frac{1}{k+\imat0^+}=\frac{\wp}{k}-\imat\pi\delta(k)
\end{equation}
allows to convert \eqref{eq:xi1xi2} in terms of the $\delta$ distribution and of the Cauchy principal value~$\wp$:
\begin{multline}\label{eq:xi1xi2bis}
  \int\xi_\kpu^*(x,t)\,\xi^{}_\kpd(x,t)\,\dmat x=\delta(k_1-k_2)\\
+\rho^*_1\delta(k_1+k_2)
+\frac{\imat}{2\pi}\left[
(\rho^{}_2-\rho^*_1)\frac{\wp}{k_2+k_1}+(\rho^*_1+\rho_2+2\rho^*_1\rho_2)\frac{\wp}{k_2-k_1}\right].
\end{multline}
In fact, the coefficients of the principal values both vanish when the respective denominators cancel
(we use \eqref{eq:rhotaumk}, \eqref{eq:1rhotau}  and \eqref{eq:energyconservation},
 then~$\rho+\rho^*+2|\rho|^2=0$ follows)
and we can drop the symbol~$\wp$. The~$\delta(k_1+k_2)$ can also be forgotten for, to constitute the basis,
 we retain only strictly
positive values of~$k_1$ and~$k_2$. A little bit of algebraic jugglery with \eqref{eq:tau} and \eqref{eq:rho}
allows to check that 
\begin{equation}
  \frac{\imat}{2\pi}\left[
\frac{\rho^{}_2-\rho^*_1}{k_2+k_1}+\frac{\rho^*_1+\rho_2+2\rho^*_1\rho_2}{k_2-k_1}\right]
=-\frac{\kappa^2}{2\pi}\,\frac{\tau_1^*\tau^{}_2}{(\Omega_\kappa^2-\omega_1^2)(\Omega_\kappa^2-\omega_2^2)}
=-\chi^*(\omega_1)\chi(\omega_2)
\end{equation}
with $\tau_n\DEF\tau(k_n)$, $\omega_n\DEF\omega(k_n)$ ($n=1,2$). Then we have  proved that, for~$k_1>0$
and~$k_2>0$,
\begin{equation}
  \Xi_{k_1}^\mathrm{in}(t)\cdot\,\Xi_{k_2}^\mathrm{in}(t)=\left(X_{k_1}^\mathrm{in}(t)\right)^*X_{k_2}^\mathrm{in}(t)+\int\xi_\kpu^*(x,t)\,\xi^{}_\kpd(x,t)\,\dmat x=\delta(k_1-k_2)\;.
\end{equation}
The space inversion of this identity leads immediately to~$\Xi_{-k_1}^\mathrm{in}(t)\cdot\,\Xi_{-k_2}^\mathrm{in}(t)
=\delta(k_1-k_2)$. At last, with the same techniques we can obtain
 \begin{multline}
 \!\!\! \int\xi_\kpu^*(x,t)\,\xi^{}_\kmd(x,t)\,\dmat x=(1+\rho^*_1)\delta(k_1+k_2)\ +\\
\frac{\imat}{2\pi}\left[\frac{2\tau^{}_2\tau^*_1-\tau^{}_2-\tau^*_1}{k_2-k_1}+\frac{\tau^{}_2-\tau^*_1}{k_2+k_1}
\right].
\end{multline}
As above, for~$k_1$ and $k_2$ both strictly positive, $\delta(k_1+k_2)$ vanishes whereas the second term on the right hand side
is precisely~$-\left(X_{k_1}^\mathrm{in}(t)\right)^*X_{-k_2}^\mathrm{in}(t)$. Therefore
\begin{equation}
  \Xi_{k_1}^\mathrm{in}(t)\cdot\,\Xi_{-k_2}^\mathrm{in}(t)=\left(X_{k_1}^\mathrm{in}(t)\right)^*X_{-k_2}^\mathrm{in}(t)+\int\xi_\kpu^*(x,t)\,\xi^{}_\kmd(x,t)\,\dmat x=0\;.
\end{equation}
The complex conjugation and the time reversal~$t\mapsto-t$ of the above relations allows to show that
the out-modes are also orthonormalised.
The orthonormalization of~$\{\Xi_{k}^{\mathrm{\pm}}\}_{k>0}$ follows from \eqref{eq:XiRXi} and \eqref{eq:R}. To sum up, we have obtained the following 
scalar products, for any~$k_1>0$ and~$k_2>0$:
\begin{equation}
 \Xi_{\pm k_1}^\mathrm{in}(t)\cdot\,\Xi_{\pm k_2}^\mathrm{in}(t)=\delta(k_1-k_2)\;;\qquad \;\Xi_{\pm k_1}^\mathrm{in}(t)\cdot\,\Xi_{\mp k_2}^\mathrm{in}(t)=0\;;
\end{equation}
\begin{equation}
 \Xi_{\pm k_1}^\mathrm{out}(t)\cdot\,\Xi_{\pm k_2}^\mathrm{out}(t)=\delta(k_1-k_2)\;;\qquad \;\Xi_{\pm k_1}^\mathrm{out}(t)\cdot\,\Xi_{\mp k_2}^\mathrm{out}(t)=0\;;
\end{equation}
\begin{equation}
  \Xi_{k_1}^\pm(t)\cdot\,\Xi_{k_2}^\pm(t)=\delta(k_2-k_2)\;;\qquad \Xi_{k_1}^\pm(t)\cdot\,\Xi_{k_2}^\mp(t)=0\;,
\end{equation}
and, when~$\omega_0>\Omega_0$ for a unique bounded state to exist,
\begin{equation}
  \Xi_b(t)\cdot\,\Xi_b(t)=1\;;
\end{equation}
\begin{equation}
\Xi_{\pm k_1}^\mathrm{in}(t)\cdot\,\Xi_b(t)=0\;;
\quad \Xi_{\pm k_1}^\mathrm{out}(t)\cdot\,\Xi_b(t)=0\;;
\quad \Xi_{k_1}^\pm(t)\cdot\,\Xi_b(t)=0\;.
\end{equation}
Three eigenbases for the Hamiltonian have been chosen: $\{\Xi_b(t)\}\cup\{\Xi^\mathrm{in}_{k}(t)\}_{k\in\RR}$,
$\{\Xi_b(t)\}\cup\{\Xi^\mathrm{out}_{k}(t)\}_{k\in\RR}$ and
 $\{\Xi_b(t)\}\cup\{\Xi^\pm_{k}(t)\}_{k>0}$. The passage from one to the other is done with unitary matrices made of
independent $2\times2$ unitary blocks of~$S(\omega)$ or~$R(\omega)$ given by~\eqref{eq:Srhotau} and~\eqref{eq:R}
respectively.


\end{document}